# Quantized conductance in split gate superconducting quantum point contacts with InGaAs semiconducting two-dimensional electron systems

Kaveh Delfanazari[1,2,3] *, Jiahui Li[3], Yusheng Xiong[1], Peng Ma[3], Reuben K. Puddy[3], Teng Yi[3], Ian Farrer[4], Sachio Komori[5,6], Jason W. A. Robinson[5], Llorenc Serra[7], David A. Ritchie[3], Michael J. Kelly[2,3], Hannah J. Joyce[2], and Charles G. Smith[3]

[1] Electronics and Nanoscale Engineering Division, James Watt School of Engineering, University of Glasgow, Glasgow G12 8QQ, UK

[2] Electrical Engineering Division, Engineering Department, University of Cambridge, Cambridge CB3 0FA, UK

[3] Department of Physics, Cavendish Laboratory, University of Cambridge, Cambridge CB3 0HE, UK

[4] Department of Electronic and Electrical Engineering, University of Sheffield, Mappin Street, Sheffield, S1 3JD, UK

[5] Department of Materials Science & Metallurgy, University of Cambridge, Cambridge CB3 0FS, UK

[6] Department of Physics, Nagoya University, Furo-cho, Chikusa-ku, Nagoya 464-8602, Japan

[7] IFISC (UIB-CSIC) and Physics Department, University of the Balearic Islands, E-07122 Palma, Spain

*Corresponding author: kaveh.delfanazari@glasgow.ac.uk   Dated: 07062023

**Quantum point contact or QPC- a constriction in a semiconducting two-dimensional (2D) electron system with a quantized conductance- has been found as the building block of novel spintronic, and topological electronic circuits. They can also be used as readout electronic, charge sensor or switch in quantum nanocircuits. A short and impurity-free constriction with superconducting contacts is a Cooper pairs QPC analogue known as superconducting quantum point contact (SQPC). The technological development of such quantum devices has been prolonged due to the challenges of maintaining their geometrical requirement and near-unity superconductor-semiconductor interface transparency. Here, we develop advanced nanofabrication, material and device engineering techniques and report on an innovative realisation of nanoscale SQPC arrays with split gate technology in semiconducting 2D electron systems, exploiting the special gate tunability of the quantum wells, and report the first experimental observation of conductance quantization in hybrid InGaAs-Nb SQPCs. We observe reproducible quantized conductance at zero magnetic fields in multiple quantum nanodevices fabricated in a single chip and systematically investigate the quantum transport of SQPCs at low and high magnetic fields for their potential applications in quantum metrology, for extremely accurate voltage standards, and fault-tolerant quantum technologies.**

***Introduction:*** Quantum point contact (QPC) is a small constriction defined in a two-dimensional electron gas (2DEG) system that exhibits quantized conductance. Generally, the constriction has a width $W_c$ and a length $L_c$ both smaller than the mean free path. This particular shape, also known as the Sharvin point contact, was proposed to study the Fermi surface in a metallic sample and has since been used to study scattering [1,2].

With the advent of 2DEG heterostructure and the need to investigate quantum transport, the QPC in a GaAs/AlGaAs heterostructure was developed, followed by the prediction of the quantized conductance as multiples of the conductance quantum $G = \sum_{n=1}^{N_c} \frac{2e^2}{h}$ [3]–[9]. Since then, QPCs have become important tools for studying electron transport in condensed matter systems and have found a wide range of applications in areas such as quantum nanoelectronics. Ever since the investigations of the semiconductor-superconductor hybrid junction were initiated, a plethora of experimental and theoretical efforts have been dedicated to this field [10]. This was followed by the theoretical prediction of quantized conductance in hybrid systems (superconducting QPC, or SQPC) [11]: the conductance of the hybrid field effect SQPC is quantized, like normal QPCs; however, the plateau height gets a different value than a normal QPC with a step height of $2e^2/h$. In perfect conditions, the conductance step height of SQPCs with a single superconductor-2DEG interface is predicted to be doubled (gets a value of $4e^2/h$) due to the retroreflective property of the scattering (retro property of Andreev reflection) at the interface [11], and is related to the number of 1D subbands in their constrictions. In such a one-interface system, the normal electron excitation incident from 2DEG is reflected at their interface with a superconductor as hole excitation with identical momentum but opposite velocity. This means that every single reflected quasiparticle initially radiated from the hybrid junction will come back to it, causing the doubling of conductance in a single interface superconductor-2DEG device [12]. In the case of a smooth and impurity-free superconducting constriction of length shorter than the coherence length ($L_c <<$

$\xi_0=(\hbar v_F/2\pi\Delta_0)$), the step height in conductance at zero temperature is dependent only on the bulk superconductor energy gap $\Delta_0$ and is independent of the junction parameters [12], [13]. However, the existence of disorder and roughness in the interfaces of the real hybrid junctions, the device geometry and fabrication errors may result in the suppression and reduction of conductance values [14]. Moreover, the conductance $G$ as a function of contact width or Fermi energy of a hybrid single interface QPC has also been predicted to have plateaus at half-integer multiples of $4e^2/h$ if the superconductor is in a topologically nontrivial phase, but with usual integer multiples in the topologically trivial phase, sensitive to disorders [15]. Nevertheless, the experimental exploration of quantized conductance values in hybrid SQPCs is limited to only a few studies on InAs-based 2DEG with observed step heights that do not exactly follow the predicted theoretical values [12,16].

When a negative gate voltage is applied to a hybrid SQPC, the conductance oscillates due to the Fabry-Perot interference of quasiparticles [17]. By further applying negative gate voltage and increasing the absolute value of the gate voltage, the 2DEG underneath the split gates starts to pinch off. A one-dimensional (1D) subband will develop in the hybrid field effect devices when the constriction length $L_c$ becomes comparable to the 2DEG Fermi wavelength $\lambda_F$ [18]. The number of subbands in a QPC should follow the relation $2L_c/\lambda_F$ and can be modulated by the gate voltage. In the case of an SQPC or hybrid field effect device under such conditions, two currents will be running through the constriction: (i) the superconducting current and (ii) the normal current. To observe the quantized supercurrent in hybrid devices, junctions with nanoscale dimensions of high-quality interfaces are needed as the magnitude of the quantized current is predicted to depend on junction parameters and barrier strength [12], [18].

Here, we report on a novel realisation of quantum nanoelectronics circuits architecture by integration of an array of hybrid split gate InGaAs-Nb SQPCs in a single chip. Moreover, we show that robust quantized conductance can be formed when the split gate voltages of the

SQPCs are swept to negative values. Finally, we discuss the response of quantized conductance to perpendicular magnetic fields for single and double-interface SQPCs.

*Wafer growth, and quantum well simulation:* The wafer was grown by Molecular Beam Epitaxy (MBE), and the layered structure is shown in the bottom half of Fig. 1(a), in which the *x* direction is the growth direction. The 2DEG was grown on a 500μm GaAs substrate started with a GaAs/AlAs/GaAs (50/70/250nm) buffer layer. On top of the buffer layer are a 1300nm linear graded InAlAs buffer layer, another 250 InAlAs buffer layer, and a 30nm $In_{0.75}Ga_{0.25}As$ quantum well which is a 2DEG with mobility $\mu_e = 2.5 \times 10^5 (cm^2 V^{-1} s^{-1})$ and electron density $n_s = 2.24 \times 10^{11} (cm^{-2})$ in the dark. The 2DEG is buried under a 60nm $In_{0.75}Al_{0.25}As$ spacer, a 15nm $In_{0.75}Al_{0.25}As$ with n-type doped modulation and another 45nm $In_{0.75}Al_{0.25}As$ spacer with a final 2nm InGaAs cap to prevent oxidisation. Figure 1 shows the wafer's simulated conduction band-edge using Nextnano (self-consistent Poisson-Schrodinger solver) [19]. The band edge is calculated from the self-consistent Poisson equation.

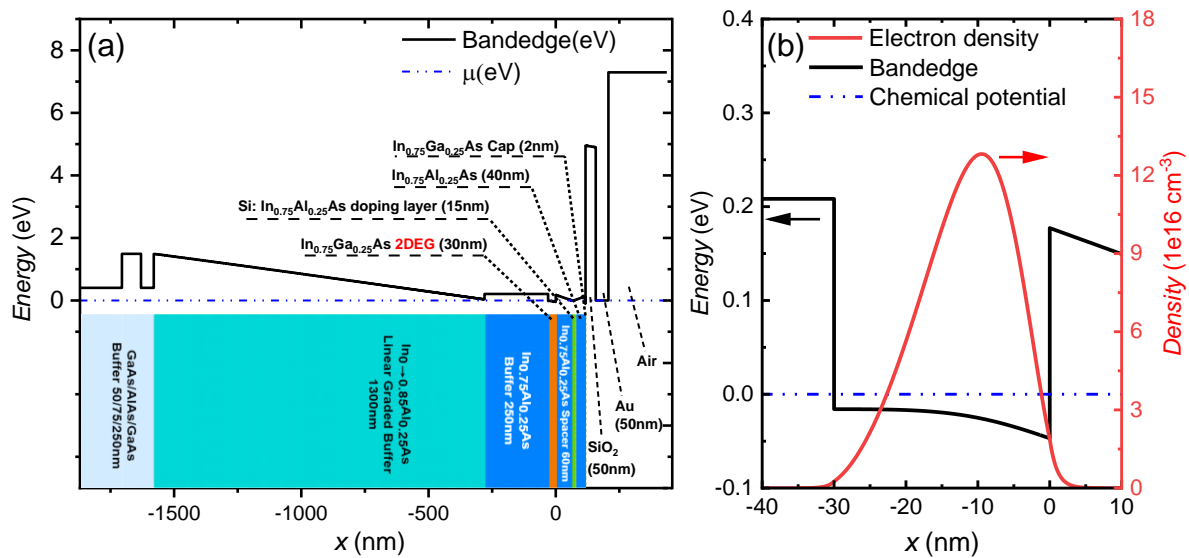

**Figure 1.** (a) Band-edge of the InGaAs wafer layered structure calculated by using Nextnano. (b) Zoomed view of the InGaAs/InAlAs quantum well. The red curve is the electron density.

The electrostatic potential is evaluated through the Poisson equation, and subsequently fed to the Schrodinger equation. The resulting redistribution of electron density will further alter the potential. This iterative process is repeated until it meets convergency requirements and achieves self-consistency. The final output is the band edges considering strain and quantum mechanics. Figure 1(b) shows the zoomed view of the 2DEG region, which is the band edge under the Fermi level. The InGaAs 2DEG layer is sandwiched between two InAlAs layers with a greater band gap. Hence the quantum well is formed in the InGaAs layer. Furthermore, the n-type silicon doped modulation layer mentioned in the wafer description will tile the bandgap towards the lower energy and bring the conduction band-edge down so that the quantum well bottom is lower than the Fermi level. A triangular quantum well is formed below the Fermi level. In this sense, the 2DEG is formed with several subbands filled with electrons. The red curve in Fig. 1 (b) is the electron density calculated with Nextnano. The integrated electron density is around $2.1 \times 10^{11} cm^{-3}$, consistent with the previously reported result [20].

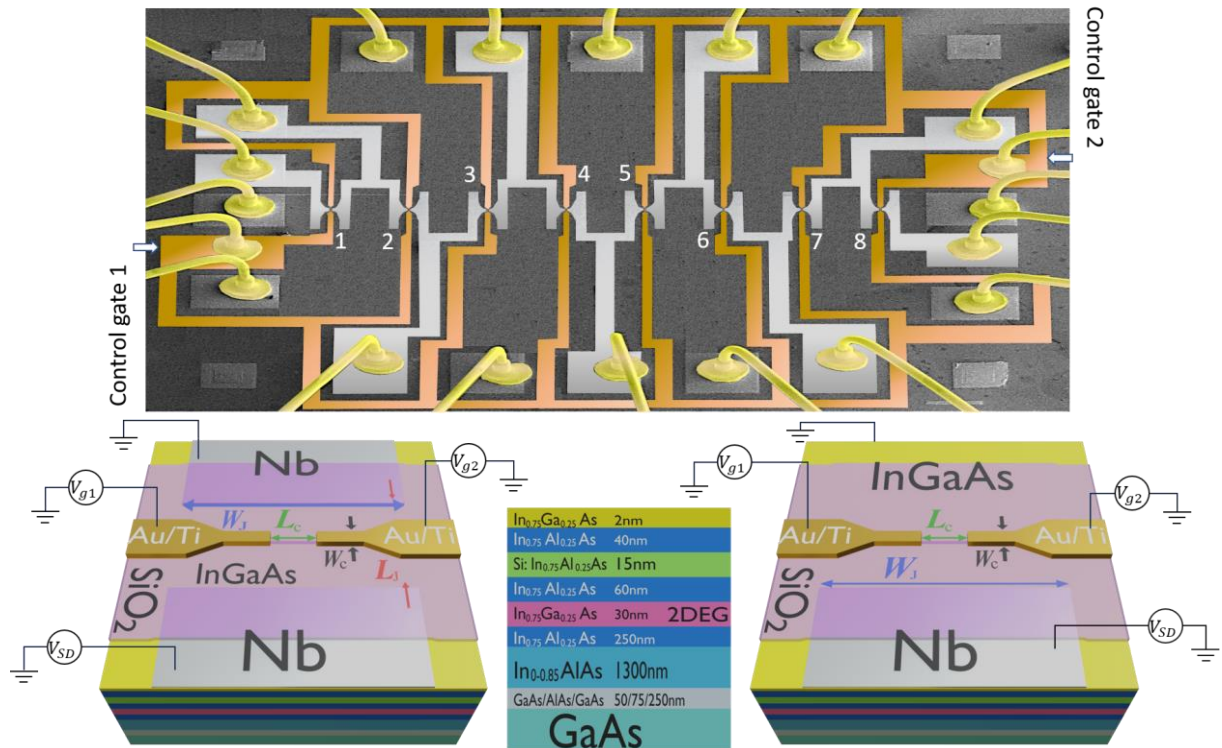

**Figure 2.** (Top) The false-colour scanning electron microscope (SEM) image of the integrated hybrid SQPC chip with eight split gate devices each controlled with two universal control gates (left and right). The Au leads are coloured with bright yellow, the Ti/Au gates with copper and the Nb layer by silver colour. On the bottom, the schematic demonstration of split gate SQPC double interface formed on an InGaAs heterostructure wafer with circuit measurement configuration (left) and a single interface SQPC (right). SQPC geometrical parameters for the junction with width $W_J$ (dark blue) and length $L_J$ (dark red), and constriction width $W_c$ (black) and length $L_c$ (dark green) for an array of devices are shown in Table 1. Middle is the heterostructure wafer showing that the $In_{0.75}Ga_{0.25}As$ 2DEG with 30 nm thickness is located ~120 nm below the surface.

***Hybrid SQPC chip fabrication, and sub-Kelvin cryogenic measurements:*** Figure 2 at the top shows the false-colour and scaled scanning electron microscope (SEM) image of the integrated

SQPC chip with eight split gate devices (see Table 1 for their specific dimensions) each controlled with two universal control gates (1, in the left and 2 in the right). The bottom shows the schematic of the hybrid SQPCs in the InGaAs 2DEG system with information about their dimensions, semiconductor heterostructure and the measurements of electrical circuit configurations. The detailed fabrication and cryogenic electrical characterisation of large-scale hybrid circuits are discussed in our recent work [21].

The hybrid SQPCs are biased with a source-drain voltage $V_{sd}$ and the split gates are separately controlled via voltage signals $V_{g1}$ and $V_{g2}$. We designed our devices in a way to have specific parameters to investigate their conductance reproducibility and robustness for their suitability for future unconventional and topological superconductivity investigation; therefore, our focus in the present study is on the conductance properties of the SQPCs devices [22]–[27]. The split gate SQPC parameters are labelled on the device surface in Fig. 2 with $L_c/W_c$ as constriction length/width, and $L_J/W_J$ as the junction length/width. For a single interface device (bottom right), there is an ohmic contact on the other side of the Nb contact.

**Table 1.** The designed geometrical parameters of eight hybrid split gate SQPCs integrated into a single quantum chip.

| SQPC parameters | 1 | 2 | 3 | 4 | 5 | 6 | 7 | 8 |
|---|---|---|---|---|---|---|---|---|
| $L_c$ (nm) | 400 | 400 | 400 | 400 | 400 | 400 | 400 | 400 |
| $W_c$ (nm) | 400 | 300 | 200 | 100 | 100 | 100 | 100 | 100 |
| $L_J$ (µm) | 1.4 | 1.4 | 1.4 | 1.4 | 1.4 | 1.4 | 1.4 | 3.2 |
| $W_J$ (µm) | 5 | 5 | 5 | 5 | 5 | 5 | 5 | 5 |

The experiment was performed in an Oxford $^3He$ cryostat with a base temperature of $T=$ 280 mK. Figure 3 plots the measured conductance as a function of split gate voltage at $B=$ 0 T for eight double-interface SQPCs integrated into a single chip. We observed stepwise changes (quantization of conductance) in the conductance of seven out of eight SQPCs on a single chip as a function of gate voltage sweep from zero to negative voltages, here to $V_g$= -1 V shown with *L*, and back to 0 V shown with *R* in the legend. For example, clear multiple conductance plateaus were measured in SQPC5, and its enlarged view is shown in the left panel of Fig. 3. Note that the contact resistance of the presented devices is larger compared with our previous studies in hybrid superconductor-semiconductor devices [20], [28], [29].

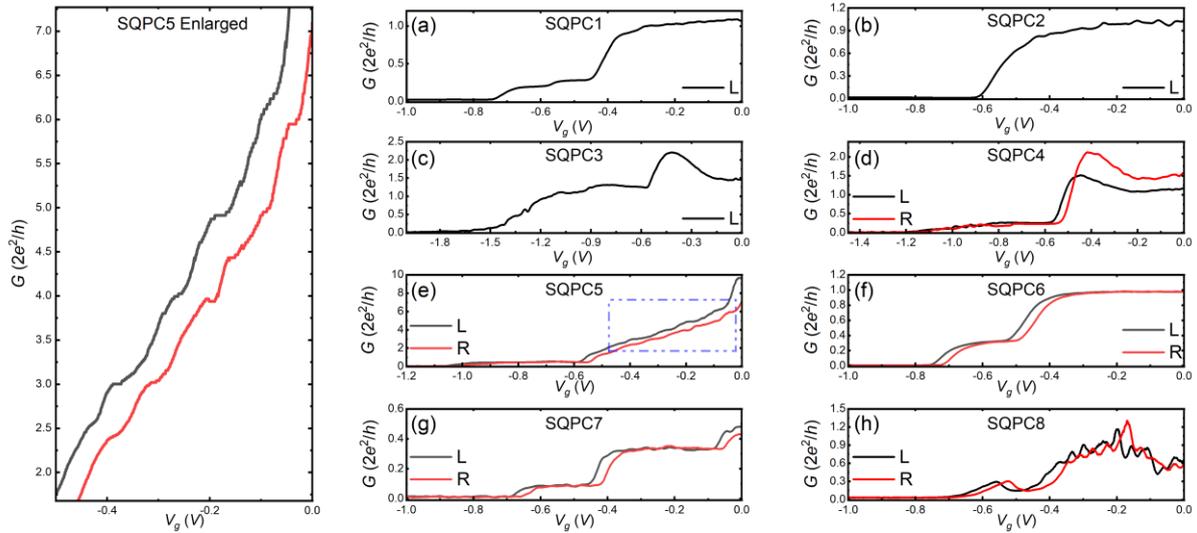

**Figure 3.** The conductance as a function of electric field on the split gates for eight hybrid SQPCs integrated into a single chip (a) SQPC1, (b) SQPC2, (c) SQPC3, (d) SQPC4, (e) SQPC5, (f) SQPC6, (g) SQPC7, and (h) SQPC8, measured at *T*= 280 mK, and *B*= 0 T. The left panel is the enlarged view of the dashed line area in SQPC5 panel, showing multiple plateaus. *L* and *R* in the legend corresponds to gate voltage sweep directions from $V_g$= 0 to -1 V and back to 0, respectively.

There are a couple of facts that affected the quantum transport measurements of our SQPCs arrays compared to a single normal QPC on a single chip: (i) our devices are 2D planar junctions and are larger, in width, length and constriction length $L_c$ than conventional QPCs [8], [28]. (ii) we etched down to the 2DEG area to make contact between 2DEG quantum wells (QWs) and Nb (normal QPCs are usually formed on the surface of wafers with ohmic annealed to melt down to the 2DEG). Some parts of QWs may be damaged or suppressed during the wet etch nanofabrication processes. (iii) there might be a different etch depth and, therefore, different heights on each side of the junctions, so the transport might be asymmetric compared to normal QPCs where a single narrow constriction of less than 100 nm is formed. (iv) there might be rough edges of the junction area underneath the split gates due to nanofabrication errors. (v) as our devices' operating temperature is at $T= 280$ mK, the observed conductance steps may be smeared out as the thermal energy, and the energy separation of the modes becomes comparable [3], [30]. A few SQPC devices show resonance structures in their conductance steps (e.g., SQPCs 3, 4, and 8), likely due to quantum interference of quasiparticles [12]–[14], [18].

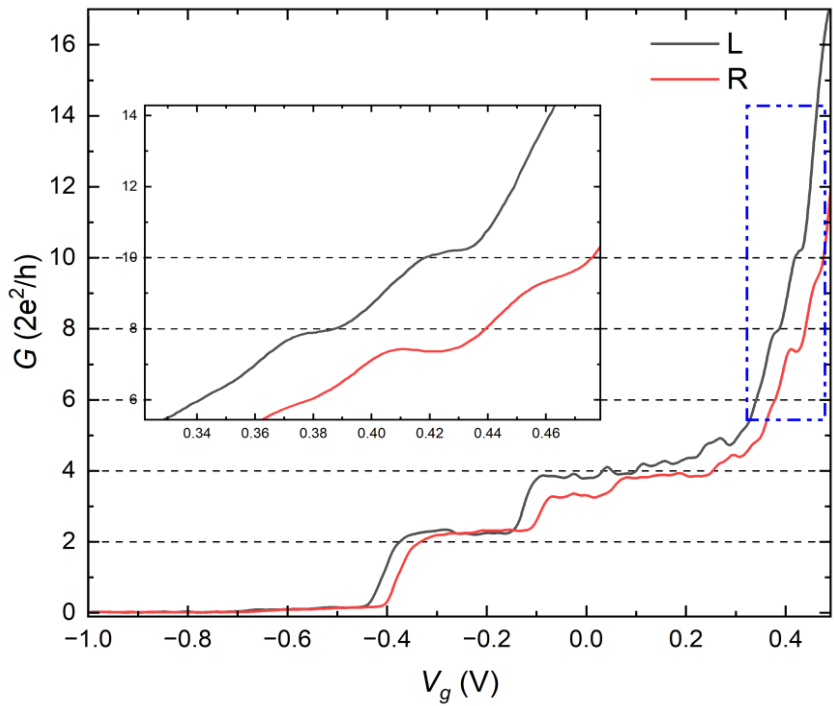

**Figure 4.** The quantized conductance as a function of the electric field under the split gates for an Nb-In$_{0.75}$Ga$_{0.25}$As QWs split gate hybrid SQPC for two sweep directions from right to left (*L*) and from left to right (*R*). It is clearly seen that quantized conductance gets values of $2 \times 2e^2/h$ for four observed plateaus for both gate voltage sweeps directions, confirming the high-quality interface formation between superconducting Nb and In$_{0.75}$Ga$_{0.25}$As QWs in the junction. The last two plateaus observed in the quantum transport measurements at $T= 280$ mK, and $B=0$ T are enlarged, and plotted in the inset.

Figure 4 plots the measured conductance as a function of split gate voltage for a single interface hybrid SQPC at $B= 0$ T. We observe a clear quantization of conductance as a function of split gate electric field for two left (*L*), and right (*R*) sweep directions. We observe four plateaus with step heights around $2 \times 2e^2/h$. The plateau $6 \times 2e^2/h$ is slightly unclear. This could have different reasons and requires further in-depth investigations. One could be due to the added characteristic of operation between the regions of positive and negative polarities for *V*g. We further studied quantized conductance behaviour in our SQPC devices by performing magnetic field-dependent conductance measurements for hybrid field effect devices with either one interface (Nb-2DEG) or two interfaces (Nb- 2DEG-Nb). The conductance of the hybrid field effect SQPC devices is susceptible to magnetic fields [30,31]. The effect of the magnetic field on their conductance at superconducting states depends on the value of absolute gate voltage. When the gate voltage is relatively large, the current path will narrow depending on the constriction size. Still, if this path is within the dimension of the order of $\lambda_F$, the conductance may be less or insensitive to small magnetic fields.

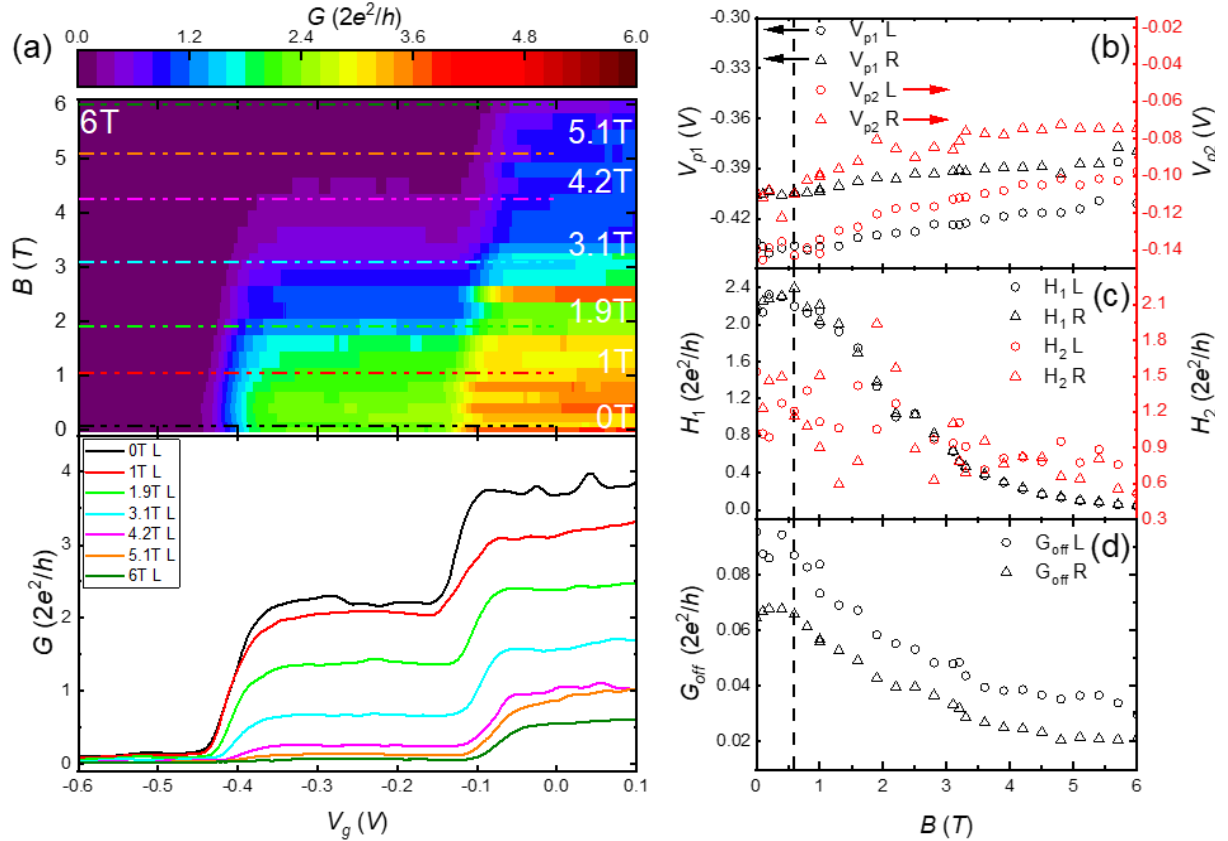

**Figure 5.** (a) The colour-coded graph on the top panel shows the conductance as a function of electric field on the split gates for a single interface Nb-2DEG SQPC hybrid device at different perpendicular magnetic fields up to $B$= 6T, at $T$= 280 mK. The line cuts are shown on the bottom panel. There is only one sweep direction from 0 to -1 V (denoted as $L$) shown and the sweep from -1 V to 0 (denoted as $R$) is shown in the appendix section. A relatively good reproducibility rate is observed for all magnetic field strengths in this device. (b) The evolution of conductance first and second steps as a function of applied perpendicular magnetic fields for two gate voltage sweep directions. (c) The conductance height for the first and second steps as a function of applied perpendicular magnetic fields. (d) The off-state conductance of the device as a function of the applied magnetic field for two gate voltage sweep directions. The dashed line is the border between the areas with a robust response against the external magnetic fields (left) and the area with decreasing quantum conductance against the applied magnetic fields (right).

Figure 5 shows the conductance versus gate voltage sweeps of a hybrid field effect SQPC with one Nb-2DEG interface at different perpendicular magnetic fields (see the caption for detailed information). The analysis in Fig. 5 (b)-(d) was extracted from the conductance plateaus on the left, in which $V_{p1}$ and $V_{p2}$ are the pinch-off voltages of the first and second plateaus, respectively. $H_1$ and $H_2$ are the conductance heights measured from the first and second plateaus, and $G_{off}$ is calculated from the average conductance when the hybrid SQPC is fully pinched off. We observe quantized conductance with step heights as large as $\cong 2 \times 2e^2/h$ at zero magnetic fields which is quite robust against perpendicular magnetic fields of up to $B=$ 0.6 T. A monotonic decrease of conductance due to suppression of Andreev reflection probability, as well as superconducting properties of Nb, is observed as the field strength increases to above $B=$ 0.6 T (see Fig. 5 (c)). Above this point, the field magnitude is high enough to suppress the retro property of Andreev reflections, so the conductance height starts to decrease gradually. At around $B=$ 1.7 T, the device response becomes almost like normal QPCs. The pinch-off voltage for the first and second steps are plotted in Fig. 5 (b), which shows a monotonic shift to higher fields (with relatively good reproducibility) for both directions of the split gate voltage sweeps. The switching point where the hybrid SQPC switches off from superconducting to an insulating state also moves to higher fields, as shown in Fig. 5 (d).

Figure 6 presents the same measurements discussed in Fig. 5 but for a hybrid split gate SQPC device with two interfaces (Nb-2DEG-Nb). In this case, the induced superconductivity and Josephson coupling are dependent on the distance between the two Nb contacts on each side of the $In_{0.75}Ga_{0.25}As$ 2DEG in the hybrid SQPCs. In case there will be no Josephson coupling for long junctions ($L_j \gg$ coherence length $\xi$), an Nb-2DEG-Nb device can be considered as two Nb-2DEG devices in series, with a conductance $G_{Nb-2DEG-Nb} \approx 1/2\ G_{Nb-2DEG}$ [32]. In this SQPC, the quantized conductance with step heights as large as $\cong 0.32 \times 2e^2/h$, for the first plateau, and $0.65 \times 2e^2/h$, for the second plateau, at zero magnetic fields are observed. The pinch-off voltage shift observed in double interface SQPCs is much more pronounced compared to that for a single interface SQPC, which might be due to the magnetic field weakening induced superconductivity or the Josephson coupling between two back-to-back $In_{0.75}Ga_{0.25}As$ QWs junctions [32-34]. As the magnetic field increases, the pinch-off voltage decreases due to weaker coupling between the two interfaces of the hybrid junction. The step height expands to $\cong 0.35 \times 2e^2/h$ and $0.69 \times 2e^2/h$, as the external magnetic field increases to $B= 0.6$ T, where the conductance starts to decrease. Current enhancement in one-dimensional hybrid superconducting-semiconducting junctions has been observed and interpreted as both nontopological [32] and topological [35] origin, but the actual cause of the effect is still under debate.

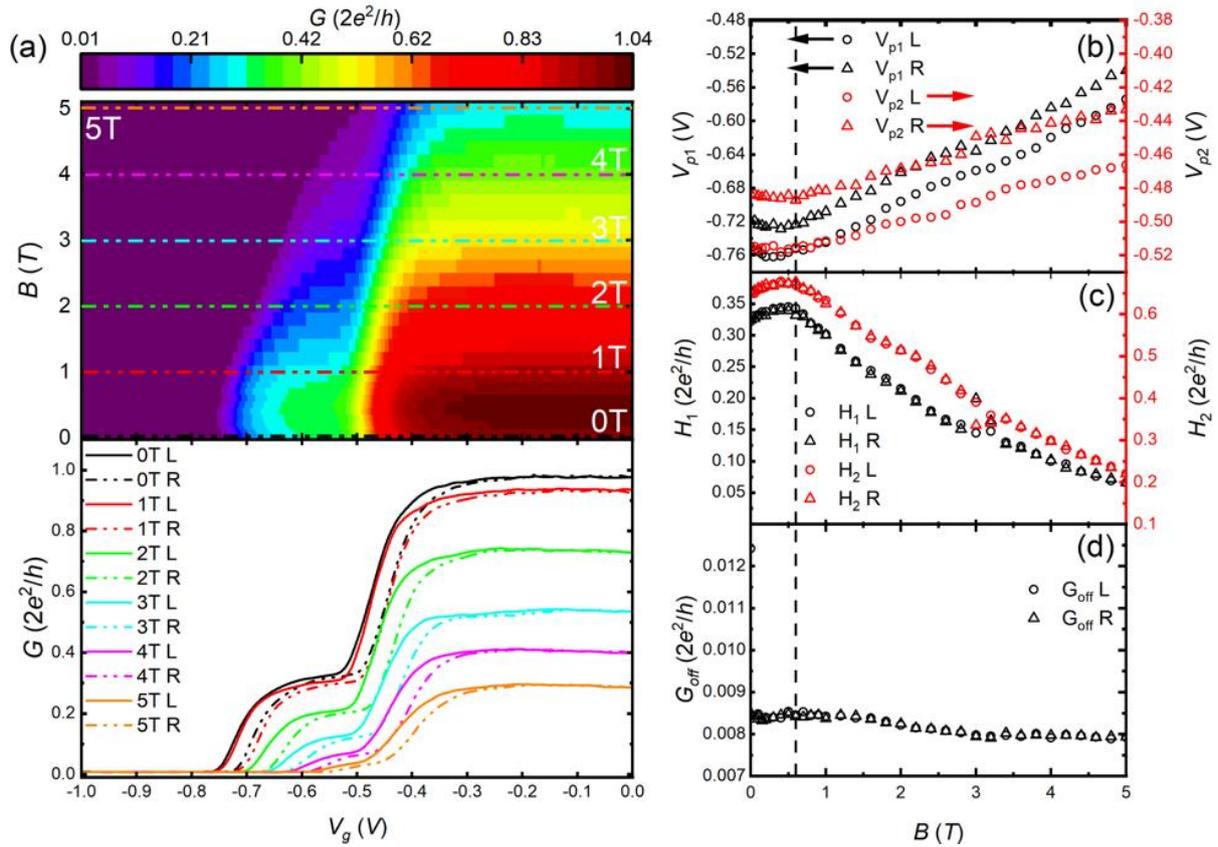

**Figure 6.** The conductance as a function of electric field on the split gates for a Nb-2DEG-Nb hybrid SQPC at different perpendicular magnetic fields up to 6 T, and at $T= 280$ mK shown as colour-coded graph with the same analysis discussed in the caption of Fig. 5.

To the best of our knowledge, the observation of the quantized conductance enhancement under perpendicular (out-of-plane) magnetic fields has not been observed before in any hybrid quantum system. In addition to the origins mentioned above, weak antilocalization may also contribute to this effect. Elucidating the underlying aspects contributing to the magnetic field-dependent increase of current and conductance in hybrid SQPCs is a very interesting topic and a strong motivation for future research works.

***Conclusion*:** We reported the first observation of quantized conductance in an array of hybrid split gate SQPC devices based on the In$_{0.75}$Ga$_{0.25}$As two-dimensional electron system and demonstrated a systematic experimental investigation of their quantum transports at zero and high magnetic fields at millikelvin temperatures. We observed quantized conductance doubling in hybrid field effect SQPC devices with single superconductor-semiconductor interfaces and found their robustness against external perpendicular magnetic fields up to *B*= 0.6 T. We further observed a clear transition from SQPC behaviour to normal QPC at high magnetic fields. By performing the same measurement on SQPC devices with two superconductor-semiconductor interfaces, we observed a stronger correlation between pinch-off voltage and external magnetic fields, indicating that induced superconductivity, Josephson coupling, split gate and interface quality play an important role in SQPC behaviour. Both single and double-interfaced SQPC devices present slight conductance enhancement under the application of perpendicular magnetic fields; this novel observation requires further investigation in a large array of hybrid devices so the statistics and reproducibility of the data may help to understand the topological or nontopological origin of the effect. SQPCs offer unique quantum mechanical properties, such as accurate control of Cooper pairs flow, the quantization of superconducting order parameters and the development of discrete energy levels. This allows the application of SQPCs in quantum metrology for extremely precise and stable voltage standards and for the development of cryogenic quantum nanoelectronics circuits, and processors with ultrahigh sensitivity. Moreover, the proposed hybrid quantum circuit architecture may be implemented to investigate unconventional and topological superconductivity in a large array of coupled artificial hybrid devices for their potential applications in fault-tolerant topological quantum technology.

# Appendix:

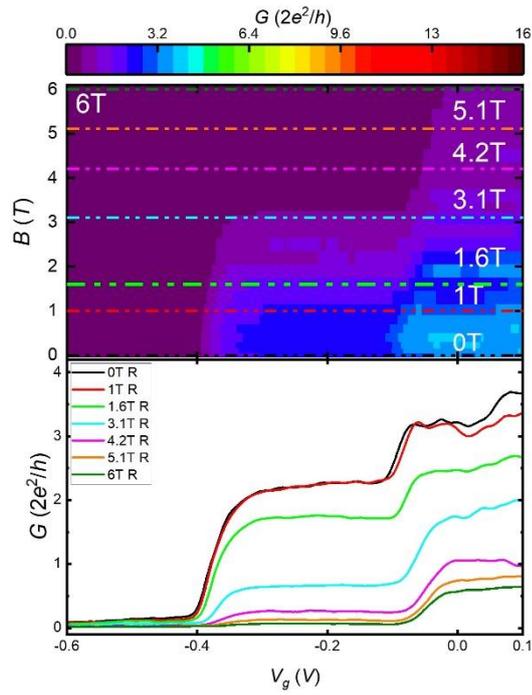

**Figure A1.** The conductance as a function of electric field on the split gates for a single interface Nb-In$_{0.75}$Ga$_{0.25}$As QWs SQPC device at different magnetic fields up to $B$= 6T and at $T$= 280 mK shown as colour-coded graph on top, where the line cuts are shown on bottom. There is only one sweep directions shown from -1 V to 0 (denoted as $R$). Comparing with Fig. 5, a relatively good reproducibility is observed for all magnetic field strengths in our hybrid SQPC nanodevices.

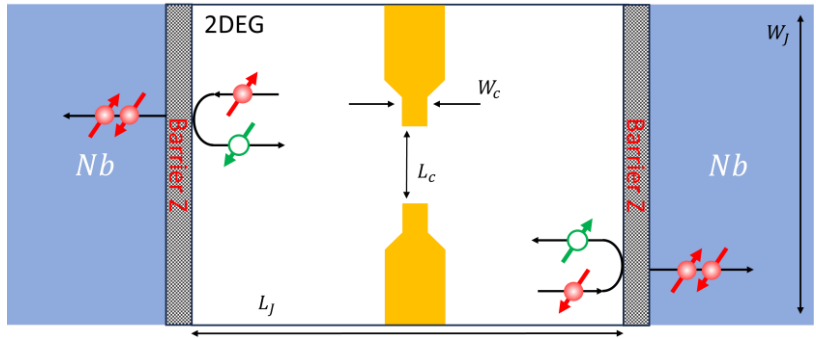

**Figure A2.** The schematic of the SQPC device structure shows shadow regions that represent possible potential barrier at the Nb-InGaAs 2DEG interfaces, with $'Z'$ represents the barrier height (strength). The Junction width $W_J$ and length $L_J$, as well as the constriction width $W_c$ and length $L_c$ are defined accordingly. Andreev reflection is demonstrated on both sides of the Nb-InGaAs 2DEG interfaces, where an electron (red) retroreflects as a hole (green), and a Cooper pair formed in the Nb side.